# Foundational guidelines for enhancing neurotechnology research and development through end-user involvement

Amparo Güemes[1]†, Tiago da Silva Costa[2,3,4]†, Tamar R. Makin[5]†

[1]Electrical Engineering Division, Department of Engineering, University of Cambridge, UK; [2]Newcastle University Translational and Clinical Research Institute, UK, [3]NIHR Newcastle Biomedical Research Centre, UK, [4]Cumbria, Northumberland, Tyne and Wear NHS Foundation Trust, UK, [5]MRC Cognition and Brain Science Unit, University of Cambridge, UK

† Authors contributed equally to this work.

**ABSTRACT**

Neurotechnologies are increasingly becoming integrated with our everyday lives, our bodies and our mental states. As the popularity and impact of neurotechnology grows, so does our responsibility to ensure we understand its particular implications on its end users, as well as broader ethical and societal implications. Enabling end-users and other stakeholders to participate in the development of neurotechnology, even at its earliest stages of conception, will help us better navigate our design around these serious considerations, and deliver more impactful technologies. There are many different terms and frameworks to articulate the concept of involving end users in the technology development lifecycle, for example: 'Public and Patient Involvement and Engagement' (PPIE), 'lived experience', 'co-design', 'co-production'. Here we utilise the PPIE framework to develop clear guidelines for implementing a robust involvement process of current and future end-users in neurotechnology, with an emphasise of patient involvement. We present best practice guidance for researchers and engineers who are interested in developing and conducting a PPI strategy for their neurotechnology. We consolidate advice from various online sources to orient individual teams (and their funders) to carve up their own approach to meaningful involvement. After an introduction that coveys the tangible and conceptual benefits of user involvement, we first guide the reader to develop a general strategy towards setting up their own process. We then help the reader map out their relevant stakeholders and provide advice on how to consider user diversity and representation. We also provide advice on how to quantify the outcomes of the engagement, as well as a check-list to ensure transparency and accountability at various stages. The overall aim is the establishment of gold-standard methodologies for ensuring that patient and public insights are at the forefront of our scientific inquiry and product development.

## 1. INTRODUCTION

Neurotechnologies are projected to increasingly become a part of our everyday lives, interwoven with our physical and mental experiences. This increased intimacy between humans and technology opens unprecedented new opportunities to improve society – but also potential risks. As neurotechnologies become more popular and their impact grows, so does our responsibility to ensure we understand their implications and the boundaries of humanity we want to preserve. Importantly, we should also be clear as a society on who gets to be involved in the decisions about how to guide the evolution of these technologies.

Neurotechnologies can be broadly defined as devices that are specifically designed to interact with the user's nervous system, by recording, extracting or modulating neural activity. This very broad definition includes neurotechnologies which can be adapted for very different classes of devices and tools, with varying applications – from entertainment to chronic symptom alleviation. This potential to interact with the nervous system in diverse applications, placing the user in the centre, is core to the promise of neurotechnologies. Successful innovation in this field should therefore incorporate the meaningful involvement of end-users in the research and development process. Collaborations between technology developers and their potential users enables responsible research practices and enriches the quality and impact of the research. This ethos is the foundation of various strategies that enable researchers, users and other stakeholders to work in partnership to guide the development of technology, and how it will be implemented. Over the last decade, researchers and funders increasingly recognize the value of public and patient involvement and engagement (PPIE). Other overlapping frameworks introduce concepts such as "PICE" (standing for 'Public Involvement and Community Engagement' or "PPCE" (standing for 'Public Participation and Community Engagement' [1]), "co-production" [2], "co-design" [3] or "lived experience" [4]. As a result, patient and public involvement activities and plans are now routinely scored by grant panels and academic journals specifically request summaries of these activities. For example, the British Medical Journal already requires all submitted manuscripts to include a "Patient and Public Involvement Statement" within the methods [5]. While 'involvement' and 'engagement' in research might seem like straightforward and overlapping concepts, in the current context they have more precise meanings [6]. Involvement refers to the active engagement of patients, the public and other stakeholders as collaborators in the design and oversight of a research project[7] The idea is that fully cooked protocols are not just presented to a group of "lay people" for tokenistic criticism but rather that those with lived experience are empowered to act as true partners. This makes research more likely to align to real-world needs and therefore more likely to be impactful. Engagement relates to broader societal reach, to increase transparency of processes and trust in outputs [8,9]. For consistency and simplicity, we made the pragmatic choice throughout this document to use the term PPI, standing nominally for 'public and patient involvement', however we note that our report and guidelines extend to neurotechnologies that are being developed for non-clinical use cases.

While the key principles of successful PPI are broadly suitable for any research, neurotechnologies may require some unique user-centred characteristics, as well as distinctive societal and ethical considerations [10,11]. Classical ethical considerations include safety and access [11–14]. Beyond these, neurotechnologies have specific ethical challenges, including how they may challenge our notions of, or even fundamentally change, human autonomy, agency, identity and responsibility, when connected to and living with a neural interface as part of our bodies. In addition, beyond traditional fairness, equity, and justice concerns, there are questions of how the use of neurotechnologies could augment human capacity. By fundamentally changing human brain function, neuroenhancement technologies could exacerbate existing inequities and divides around technology. Similarly, familiar concerns around data security and privacy may be particularly concerning in relation to brain or "mental activity": as technology evolves, and mechanistic processes are unveiled, what will we make of measures and recordings of thoughts or feelings? How will they be interpreted, and for what uses? [15]. These and other ethical questions posed by the development of neurotechnological advances propelled the need to integrate neuroethics into the engineering process [11,13,14]. For example, there is an emphasis on consent measures to empower neurotechnology users' control over their data. While medical devices will more frequently have robust safeguards built into certification processes, recreational and "well-being" devices will occupy a much less regulated space. The expanded use cases for these technologies will continue to blur the lines between clinical and non-clinical use. This highlights the recognised need to build trust and accountability in neurotechnologies. This can be enabled by better involvement of the relevant stakeholders in the research and development of new technologies, as well as broader engagement and information sharing. All of this is of particular concern in the context of an international governance vacuum, especially for non-clinical applications [see chapter 2 of the UNESCO report *Unveiling the Neurotechnology Landscape* [16]].

There are also potential benefits for researchers/developers in co-designing and implementing a successful PPI strategy. This could start with the identification of study partners/co-applicants with lived experience and could end with more effective dissemination to ensure wider impact and applicability of the findings [17,18]. PPI could benefit multiple research stages, starting from identifying research opportunities and fundraising, to co-design of technology elements, co-development of user-facing information and materials, involvement in synthesis of research findings and dissemination activities[17]. However, a PPI strategy can be challenging to plan and manage (for a good summary we recommend Section 1.2.1 in [19]).

For these reasons, many major funders are now requiring that PPI activities and their impact are embedded and formally costed in applications, being used as an assessment criterion when reviewing grants – both the PPI strategy as part of the research proposal, but also evidence that there was end-user involvement in the development of the proposal itself. Leading examples, from which the neurotechnology field could learn from and build upon, mostly come from the biomedical field. In the UK, this includes major funders, such as the National Institute for Health and Care Research (NIHR) [20], UK Research and Innovation (UKRI) [21] or the Wellcome Trust [22]. In the European Union, the Horizon Europe funding programme [23] strongly values the demonstration of Responsible Research and Innovation (RRI) principles in grant applications, including public engagement [24]. Further afield, the Canadian Institutes of Health Research [25] the National Health and Medical Research Council of Australia [26] and the United States of America National Institutes of Health [27] all offer toolkits and guidance on patient and public involvement, reflecting its increasing importance for successful applications. track record in setting up a PPI programme is becoming a valuable qualification for applicants.

In principle, the relevant stakeholders can be involved in the earliest stages of the research cycle (e.g., agenda setting, grant writing, research design). Indeed, there are concrete benefits for enabling PPI as part of the full cycle of innovation, from ideas to adoption [28]. Rather than developing a technology and testing post-hoc whether it met the needs of diverse users, it is more beneficial, efficacious, sustainable long-term and ethical to start with the identified needs of users (e.g. in mental health [29]), involve them in the development of a roadmap that also addresses user-centred concerns. Integrating continuous and bidirectional feedback from partners throughout the research process has the potential to enhance the quality, relevance, and ethical foundation of innovative research, and is expected to result in greater impact [28].

Our report does not aim to provide a comprehensive overview of past PPI case studies. Instead, it seeks to emphasize the necessity and utility of a unique framework for PPI within the context of neurotechnology development and to guide researchers on best practices and key considerations for its implementation. As a result, we propose guidelines, designed to involve potential neurotechnology users, their communities and carers, and other relevant stakeholders, regardless of their clinical state: we consider patients, other consumers / end users, and also the general public, with the aim to establish partnerships with experts by experience with neurotechnology.

## 2. PRACTICAL CONSIDERATIONS FOR PPI STRATEGY PLANNING: ADVICE FOR GETTING STARTED

We strongly recommend to pre-register the PPI strategy, including goals, desired outcome measures, activity, and makeup of the PPI prior to conducting the involvement. This will improve the transparency of the project, will allow to seek feedback at the earliest stages of the plan from peers and relevant stakeholders, and will help demonstrate your commitment to the relevant communities. It is also advised, whenever possible, to include the PPI partners in the development of the PPI strategy. As it stands, we recommend the PPI strategy is pre-registered as part of the study protocol, as we are not aware of specific repositories with any relevance for neurotech.

### 2.1 Mapping out the motivation for a PPI plan and general strategy

It is crucial to begin PPI work by considering the motivation for conducting the PPI and constructing a strategy to meet it: define aims, develop the appropriate methods to achieve them, and delineate clear milestones. There are several free online tools available to help structuring this:
- In the UK, the NIHR Newcastle BRC and Newcastle University have developed a free online comprehensive PPI planner tool, including the ability to include different project stages, download and share the plan [30].
- As part of the large Horizon 2020 grant programme, the EU funded the development of Engage2020 [31], including the Action Catalogue [32], a tool for enabling inclusive research (the scope goes beyond PPI but it remains useful).
- In the US, Emerson College offers the Public Engagement Roadmap, with guides, toolkits and case studies [33].

Regardless of which tool might fit better with your project, Table 1 present some key questions to guide PPI within a project. If you already have access to individuals with relevant lived experience, or related user advocates, you could ideally share these questions (and your answers) with them for early feedback.

| PPI aspects | Questions | Resources |
|---|---|---|
| **Purpose** | What is the purpose of setting up PPI within the project? Are you looking for involvement with the development of project proposals? Investigate potential users' attitudes toward neurotech research? Test specific features of the technology? Decide on user interfaces? Create rapport for support with dissemination activities? Promotion and dissemination of the project to aid recruitment? Or long-term oversight over project progression? | |
| **Involving stakeholders** | Who are the relevant stakeholders? What experiential expertise do you require? Who do you want to involve and how will you reach them?<br><br>What would be the potential benefits of/to the stakeholders taking part in the PPI? How might you support the development of your PPI partners?<br><br>Where in your plan could you allow for flexibility? For example, will people need | See Section 2.2 |

| | to travel, or can they be involved remotely? | |
|---|---|---|
| **PPI leadership** | Who will be the PPI lead? | This role is usually distinct from other essential aspects of PPI organisation/admin/facilitation. Instead, the lead should have the relevant skills, experience and authority to embed and be accountable for the PPI in the research programme. NIHR have details of what this involves [34]. Terminology will vary across the globe: The Australian Health Research Alliance will call this role "Consumer and Community Involvement (CCI) Coordinator" [35], in the US the Patient-Centered Outcomes Research Institute will use the term "Engagement Specialist"[36], among many other examples. |
| **Resources** | What type of resources and staff capacity / time will you have available to support your PPI strategy?<br><br>What level(s) of involvement will be appropriate for satisfying the project's aims and feasible within the budget / time available? (e.g., surveys, ad hoc focus groups, consultation workshops, advisory boards, integration in the team as co-researchers). What is the timeline? How often will input from the target PPI partners be needed? | See Sections 2.2 and 2.3. In addition, here is an example of a schedule of PPI activities developed by the NIHR Cambridge Biomedical Research Centre (BRC) [37]. |
| | What additional resources (time, cost, expertise) will you need to meet your various engagement levels? | The UK National Co-ordinating Centre for Public Engagement has a helpful list of funding opportunities available to specifically support health and life sciences PPI activities, many of relevance to neurotech [38]. Analogous opportunities in the US include the Patient-Centered Outcomes Research Institute Eugene Washington Engagement Award Program [36]. You should also remember to cost PPI costs into your "main" grants – this is perfectly legitimate and should be seen positively by funders. |
| **Shared decision making** | What are some key specific open questions/decisions that you would like help with? | If you are working with patients, NICE has helpful guides on shared decision making for clinicians that could be translatable into this space [39]. The EU funded Action Catalogue[32] should also be very helpful with this. |
| **Delivering and advertising** | Who could help you in delivering your plan? (e.g. institutional support, charities, funders, user groups).<br><br>How will you advertise the PPI | For example, if looking to set-up a focus group you might find that patient groups already have well established forums that you could be invited to. |

| | opportunities inclusively in practical terms? | |
|---|---|---|
| **Funding and incentives** | How will you access funding to pay people for their time, and help support their involvement, e.g. travel or carer costs? | Guidance on payments for PPI is available, for example, from NIHR including a cost calculator[40], and Health Data Research UK[41]. The Canadian Institutes of Health Research has guidance on this, based of the Strategy for Patient-Oriented Research (SPOR)[42]. The University of Manitoba (Canada) offers a free budgeting tool for PPI[43], in the form of an interactive Excel spreadsheet[43] which allows for the export of comprehensive one-page summaries (handy for grant applications). |
| **Evaluating PPI strategy** | How will you track PPI work to remain accountable to plans, ensuring PPI work doesn't become tokenistic? How will you use this to measure the impact of PPI on the project? | See section 2.4 |

**Table 1.** Key questions to guide each aspect for the PPI strategy

### 2.2. Mapping out the relevant stakeholders: Who to involve in the PPI plan

When developing a neurotechnology, it can be challenging to know who the research needs to engage with. While patient group are often associated with PPI initiatives, the future application space may extend beyond clinical indications, including the enhancement of cognitive or physical abilities in "healthy" members of the public. If the device is intended for the open market, the possibilities are endless. Therefore, it's important to involve potential future end-users from the general population, experts in ethics and technology law, representatives from industries likely to adopt such technologies, and even policymakers who can navigate the societal implications of widespread neurotechnology use. For a panoramic view of how the technology will fit with the abilities, needs and beliefs of the users, the PPI should reflect the full range of potential stakeholders, and not just the primary user group.

Pragmatism will be necessary. Start by defining the target group(s). This should include not only the target user or clinical community, but any other member of society that could be impacted by the technology, both in the immediate and the long term. For example, those caring for the primary users. We might also benefit from the insight of users of more mature neurotechnologies, who may share some common ethical and/or practical considerations. An efficient initial step in identifying key partners might be reaching out to disease-specific medical charities, which can provide essential resources, information, and connections to relevant stakeholders – this will of course be site specific and require networking. It is also important to consider a variety of relevant demographics, as people from different backgrounds, particularly population groups who are often underserved by research, might offer an overlooked, yet important perspective to neurotechnologies. It is often also worth approaching those who seem more likely to object/reject the technology, as they might offer a valid and constructive viewpoint to draw from. A diagram illustrating the different roles identified in the decision-making process of PPI activities is shown in Figure 1.

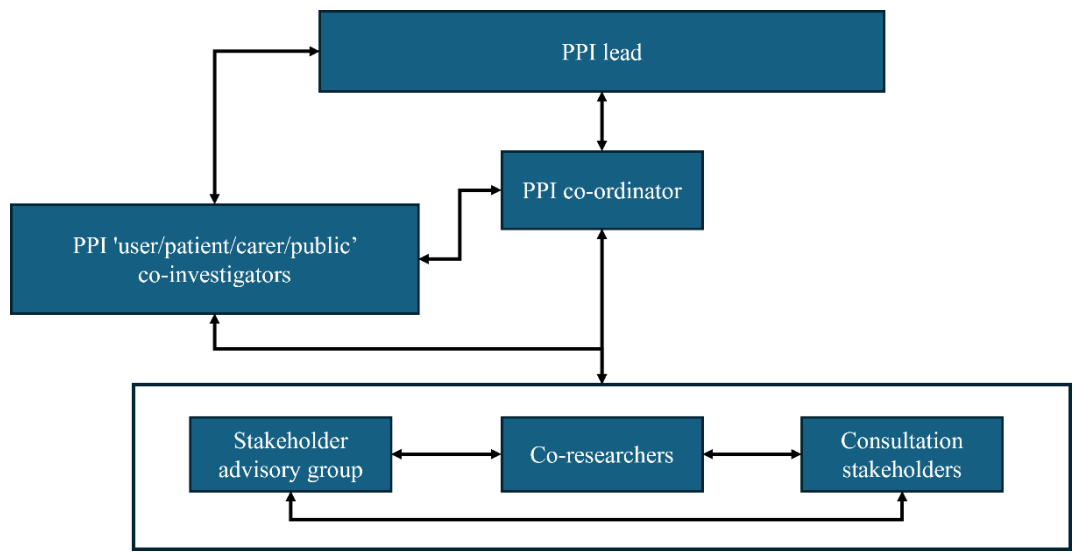

**Figure 1. Roles in PPI decision-making activities.** Identified groups work together to ensure meaningful involvement and engagement of patients, public, carers, future end-users and stakeholders throughout the research process. The PPI lead serves as the principal investigator and oversees the PPI strategy. The PPI 'User/Carer/Public' co-investigators work alongside the lead, mentored by them – not every project will require this, you must be honest about what you can offer. The PPI Coordinator is a staff member in the team who holds the PPI relationships with stakeholders and organizes PPI activities. They work closely with all other team members. The Stakeholder Advisory Group usually consists of 10-12 anchor public members who provide long-term engagement and expertise throughout the duration of the study. Co-researchers participate in specific elements of the PPI framework, such as usability testing workshops, supporting participant recruitment, or conducting user-experience qualitative work. Stakeholders from the community are invited to consultation meetings and form a network of engaged stakeholders, contributing with their perspectives and insights to the study.

Here are some general questions to ask for helping identify the desired PPI partners:
- Who are the direct users/beneficiaries of the neurotechnology under development? There are available tools in the UK to help identify populations affected by a particular condition [44]. Also, the Association of Medical Research Charities (AMRC) in the UK offers a directory [45]. This will be highly localised but it is not hard the search the web for directories (for example, in the US)[46].
- Who are the specific population groups who will be involved, and in particular those who's affiliation might influence their attitude towards the technology (e.g. religion or cultural preferences)?
- Who are the indirect beneficiaries (e.g. care-givers, healthcare professionals, employers) of successful implementation of the technology?
- What other potential user groups would be impacted from the technology in the long-term, if it is successful?
- Who will be choosing/administering the technology? Who will be paying for it? Who will commission its use in clinical groups or more widespread use? Is it all paid up-front or are there ongoing costs? Is it to be used forever? When will it need replacing?
- Who is likely to object to the implementation of the neurotechnology?
- Who is likely to be displaced / deprecated if the neurotechnology becomes a success? Will some aspects of the service they provided be missed?
- Are there any relevant technologies that are more mature in terms of societal penetration?

## 2.3. Recruiting diversity of partners required and reaching the hard-to-reach

There is a need to develop a strategy that ensures that the relationship with PPI partners is mutually-beneficial, inclusive, diverse and representative of the populations to work with. If groups are under-represented in the initial stages of technological development, then the technology might not meet their needs. There are multiple challenges. Proactive and reciprocal engagement with a broad range of views within the relevant population and ensuring they are appropriately supported to facilitate their involvement is a major consideration. As PPI is about the inclusion of individual experiences, individuals in groups will have different interests in why they are getting involved and what they are hoping to get out of it. Focusing on people with a specific demand will help understand the perceived needs and hopes for neurotech development in that particular area.

Balancing these needs will require cultural sensitivity and the engagement of a whole community, crucial for building trust. To build and maintain this trust, it is vital to recognize participants as equal partners in the process, ensuring they feel seen, heard, and understood, and to be transparent about the project's objectives and their role within it. For example, when not really offering to build a patient-led study or a co-produced study, it is strongly suggested to say so and also be clear about where the help is needed. Like all aspects of the project, the involvement strategy may have limitations, and therefore it is advisable to maintain a continuous evaluation and adaptation strategy.

To ensure that we reach the relevant partners and inclusivity, it is crucial to consider the key areas of focus described in Table 2.

| Diversity | Diversity in the target group as well as why specific kinds of diversity might be important to represent in the PPI groups. As much as possible, this should also be true of representation in PPI leads and coordinators. This may include age, gender, physical, education level, cognitive ability, race, lifestyle, religious beliefs, cultural, financial ability, societal preferences, and technological affinity. The bottom line is that there is no "general public" and we need to define the characteristics of the people that have the relevant experience for project development, while bearing in mind the target group characteristics. Also acknowledge the PPI process in itself might lead to a redefinition of the target population. |
|---|---|
| Community Engagement and Partnership Building | Leveraging national structures and building reciprocal, long-term relationships with community-based organizations, both for patients and non-patients [47]. These can be very important ways to reach some key population members (group leaders / gatekeepers / social hubs) for "snowball" recruitment. At the same time, it is important to make sure to include multiple perspectives/organisations as our partners.<br>- There are national structures that can help to identify partners. In the UK, this includes many NIHR initiatives, such as the INCLUDE project [48], the Equality Impact Assessment (EqIA) Toolkit [49], or the Health Determinants Research Collaborations [50]. In the EU, the European Patients' Academy on Therapeutic Innovation (EUPATI) [51] is a valuable resource to connect with patient groups. he Canadian Institutes of Health Research (CIHR) offers the Strategy for Patient-Oriented Research (SPOR), which has an established Alliance of public and not-for-profit agencies [52]. In Australia, the National Health and Medical Research Council (NHMRC) offers well established links to consumer and community representatives for peer review of grants [53].<br>- Community organizations also play a significant role, such as the Patient-Centered Outcomes Research Institute (PCORI[36]) in the USA, Health Nexus in Canada [54], or the Consumer Health Forum (CHF) in Australia [55]. In Asia, not for profit organisations such as Swasti can be very helpful in enabling researchers to share their PPI opportunities[56]. For non-patients, these can be members of "do- |

|  | it-yourself" communities, using readily available components, commercial products, or ad-hoc adaptations beyond the marketed intent. It might be worth searching for online discussion forums, such as those for "internet of things" builders.<br>- It is also worth considering if there are opinion makers, such as columnists, bloggers or social media influencers already talking about the technology. These might be important people to reach out to. |
|---|---|
| **Accessibility** | Shared language that is accessible. Tailoring the communication, starting with avoiding technical jargon. You might need different materials for different stakeholder groups, and even consider the need to translate it to different languages. Use of trusted services is encouraged, as data privacy will be a concern – 'the big word' is an example of service provider in this area [57]. |
|  | Accessibility in physical and digital environments and logistics. Some people might be reluctant or unable to meet online, use certain digital platforms or meet in groups in person, at least until there is shared trust. Be flexible but also honest about what to offer (and budget for). |
| **Clarity** | When reaching out to individuals or groups, we need to be clear in the aims for the PPI, as well as desired roles and expectation for the various parties involved (and commitment to agreed-upon outcomes by which such PPI efforts should be measured). There may be stages in the research cycle where PPI work may have less impact / restrictions that limit how much can be changed. It is therefore important to be open and honest in the communication on the potential and actual impact of the PPI activities at various steps of the way. |
| Incentives and benefits. | The time people spend in PPI activities should not be taken for granted. This is also particularly important if we want to reach those that incur a cost during involvement (e.g. time off work, childcare, travel expenses). It is also important to recognise that different people have different motivations when engaging in PPI activities, and some will not ask for payment, some are not able to receive payment – medically retired, or benefit status. We will need to consider alternative ways of recognising the contributions, for example providing education / training / or (access to) digital technologies. |
| Inclusive Outreach | Widening participation to encourage people to reach out to and volunteer to contribute to the PPI. Contact organisations / groups that work in the area and asking them to share your call for PPI on their social media accounts is suggested. |

*Table 2.* key areas of focus to ensure effective and inclusive engagement in PPI

**2.4. Setting out clear outcomes and success measurement for the PPI strategy:**
The PPI is meant to improve the quality of the research and successful adoption of the neurotechnology. To build evidence on the impact of the PPI activities, justify the allocation of resources, and better plan for future projects, PPI should be formally reviewed.

After setting out general motivation and key stakeholders, the next step is to clearly outline the desired outcomes and methods to get there. This could be achieved, for example, by identifying what are some potential considerations that might impact users' attitudes towards the technology (e.g. astatic, practical, societal, privacy and data security). Here again, it is important to consider the unique concerns that neurotechnologies introduce, including questions concerning autonomy, agency, identity and responsibility for a brain in a neural interface; societal consequences of widespread technological applications; control of neural data and data privacy, and access to technology and other inequalities ensuing non-inclusive design.

Ensuring that the objectives are specific, measurable, achievable, relevant, and time-bound (SMART) is essential. As much as possible, we recommend using standardised measures.

With regards to **PPI metrics**, here are some standards of good practice:
- Documenting all PPI activities to be able to report on them and identify specifically what value each opportunity brought. Some guidance documents will call this an "impact log" [58].
- Designing qualitative feedback surveys and questionnaires to identify points of consensus and diversity in values and preferences associated with the PPI by all those involved in the project [59].
- Asking people to reflect and provide feedback during semi-structured interviews of representatives with relevant lived experience about their perspective and recommendations for the neurotechnology under development [59,60].
- Feedback forms for individuals involved in the PPI to record values and preferences [61,62].
- Metrics with descriptive statistics of PPI participants key demographics that help to evaluate the diversity aims, as stated in the previous section.
- White-papers, consensus reports or recommendation reviews, co-authored by the PPI stakeholders or including contributions from the PPI group.
- Successful funding or publications facilitated directly by the PPI strategy and teamwork.
- Sharing your PPI materials and outcomes with other relevant parties.
- If the aim was to co-produce the study, then assessing how power was effectively shared.
- There is also value in subjective measures, e.g. by reflective processes. This could help identifying how the people involved found the experience, and whether they experienced any benefits from the engagement.

With regards to the **PPI impact** on the overall project outcomes, it is important to be specific about what benefits and changes it brought to the project, particularly when reporting to grant panels. Additionally, it is important to consider the need to offer feedback to those that participated in the PPI activities. In the UK, NIHR has specific guidance on this [63]. In the US, PCORI also offers specific guidance on ongoing review and assessment of engagement, including a case study [36]. A clear goal should be set out in advance for the PPI activity against which outcomes can be evaluated. This could involve:
- Ways in which PPI helped identify new values to influence the research and development strategy, e.g. incorporating user testing at an earlier phase of the process. Did it change the research priority?
- Improving the potential feasibility of the technology, e.g. by raising real-world practical considerations that might have been overlooked by the research team. Did it change the design in any way?
- Contributing to broadening the inclusivity of the neurotechnology, e.g. by addressing new diversity considerations by the team.
- Helping change people's attitudes towards neurotechnology, e.g. by convincing a grant panel that the technology will be well received by (or better tailored for) the intended users.
- Improving the resources of the research project, e.g. by increasing the diversity of the participant's pool.
- Informing strategies for managing personal data, e.g. by creating new choices for users on how their data should be collected, processed, shared, stored or protected. This should include consideration of the mechanisms for "option in" of the sharing of neural data, and what this means in the context of the proposed project.
- Identifying new factors that might enable or inhibit the research strategy, e.g. by realising diverging preferences by different user groups.
- Benefiting the involved individuals, e.g. by building trust around the use of neurotechnology.

- Contributing to the confidence and the ability of the researchers to conduct successful PPI.

## 3. Improving transparency and accountability - Recommended practices

Table 3 provides a detailed overview of these practices, offering actionable steps that can be implemented to foster greater transparency and ensure accountability at all levels.

| PPI Priority | Strategic Area | Inclusion Actions |
|---|---|---|
| **Engagement and communication** | **Bidirectional communication** | PPI involves bidirectional input, where active listening, responsiveness, and mutual understanding are key to developing effective and patient-centred neurotechnology solutions. |
| | **Long-term engagement** | This is particularly important in neurotechnology because the research often extends over extended periods. Planning for sustained PPI activities and encouraging long-term bidirectional commitment and retention of PPI representatives throughout the lifecycle of a study or project is crucial to maintain continuity, relevance, and impact. Sometimes it may be challenging to implement long-term involvement due to restrictions associated with the grant-related nature of the funding for PPI. In this case, one option is to consider providing community groups or patient advocates with access to university facilities, or offering them training and capacity-building programs that empower them to take on more active roles in the research process, contributing their skills and expertise. |
| | **Early Engagement** | Identify and involve relevant people from the earliest possible stage in the research process, ideally from shaping the initial research question and PPI strategy. |
| | **Continuous Engagement** | Plan and budget for additional involvement throughout the study, highlighting the value of bringing in fresh perspectives at key stages. |
| | **PPI Strategy Onboading** | Onboard and induct the PPI strategy from the beginning: explain the goals, action plan, timeline, and identify any gaps, so PPI partners have clear expectations and understand their roles. |
| | **Continued Partnership Opportunities** | Discuss opportunities to continue partnership in later stages of research, like PPI evaluation and dissemination activities. |
| | **Outreach and Education** | Conduct community outreach programs, such as educational sessions, workshops, virtual reality demonstrations, and webinars, to raise awareness about neurotechnology, its potential benefits, and opportunities for involvement. Be mindful of the costs of these activities and digital skills to ensure inclusivity. |
| | **Collaboration and Feedback** | Collaborate with identified stakeholders to create patient/user journey maps, identify pain points, preferences, and unmet needs that can guide the development of targeted neurotechnology solutions. |
| | **Prototype Interaction** | Allow individuals to physically interact with early-stage neurotechnology prototypes, enabling them to provide feedback on comfort, fit, and usability. Be open to adjusting design elements based on this feedback. |
| | **Information Accessibility** | Offer plain language summaries, infographics, and video content to enhance understanding of complex neurotechnological concepts. |

| Acknowledgment | Incentive Recognition | Acknowledge the motivating incentives of each of your partners. While some may gain inherent value from involvement, others may seek opportunities for additional training to gain professional skills. |
|---|---|---|
| | Unique Experiences Acknowledgment | Explicitly acknowledge the unique experiences and insights of the target users, relatives, and caregivers in the development of neurotechnological solutions. |
| | Integration of Contributions | Ensure that all contributions are highlighted and integrated into the development process. Provide regular updates on how their insights have directly impacted the design and functionalities of the neurotechnology being developed. |
| | Feedback Consolidation and Reporting | Consolidate feedback based on success measures and report it back to PPI partners. This can also guide them to tailor their support. |
| | Insight Presentation | Present insights in a clear and accessible manner, demonstrating how their participation informed decision-making throughout the development process. |
| Trust and Transparency | Trust Building | Build trust by involving individuals with relevant lived experience in supporting the design and overseeing the study from its earliest stages, demonstrating a commitment to incorporating their feedback and concerns. |
| | Transparent Documentation | Clearly document the neurotechnology development process, including the role of the PPI partners and public input at various stages, in publicly accessible reports and websites. |
| | Ethical Use Strategy | Be clear on the strategy for an ethical use of neurotechnologies in the PPI plan. PPI partners should also have a say in shaping guidelines regarding the ethical use of these technologies. |
| | Risk Identification and Mitigation | Involve PPI representatives in identifying, understanding, and mitigating risks associated with neurotechnologies, including physical, psychological, or privacy-related risks, and ensure transparency on them. |
| | Decision Sharing and IP Management | Publicly share decisions related to the design, development, and delivery of neurotechnological solutions, clearly explaining the reasoning behind each decision. Engage in transparent discussions regarding intellectual property rights and address concerns about pricing, availability, and equitable access. |
| | | Addressing concerns about the influence of IP rights on pricing, availability, and equitable access to neurotechnological solutions. |
| | | Provide a platform for stakeholders to voice concerns about any decisions that may impact the neurotechnology's accessibility or usability. |

**Table 3.** Recommended practices for improving transparency and accountability

**CONCLUSION**

The field of neurotechnology stands on the brink of transformative developments that promise to redefine our interaction with technology. From treating neurological disorders to enhancing individuals' wellbeing, the potential applications are vast and deeply impactful. Embracing a well-structured end-user involvement process, such as PPI approach, maximizes the benefits of these advancements. By integrating end-users' insights from the conception phase, neurotechnologies are tailored to meet actual needs and contexts, enhancing their usability and relevance. This user-centric approach ensures that technologies are not only more likely to be adopted but also more effective in their application, and are more likely to be trusted by the public. By addressing real-world considerations and preferences, PPI helps bridge the gap between theoretical design and practical utility, making technologies both more used and more useful.

Although PPI strategies are inherently diverse, reflecting the specific goals and challenges of each neurotechnology device, they share common objectives that benefit from shared standards and cross-project learning. Developing a gold standard for PPI is essential not only to prevent tokenistic practices but also to ensure that the involvement is meaningful and impactful. Establishing these standards encourages consistency and quality in PPI efforts, which in turn facilitates the evaluation and improvement of PPI practices. This is particularly important in a field as dynamic and ethically charged as neurotechnology, where the stakes of engagement are high due to the profound implications on users' autonomy and privacy. By exchanging knowledge and experiences through gold standards, our community is better positioned to achieve collaborative learning towards a richer understanding of how to engage effectively with different populations, including those that may be underrepresented or harder to reach in traditional research paradigms. As such, the current guidelines are likely to continue to develop, as we refine our communal understanding of the unique considerations involving neurotechnology. We hope that our foundational guidelines will help enhance the full potential of neurotechnologies in a way that is ethically responsible and socially beneficial.

**ACKNOWLEDGEMENTS**
All authors gratefully acknowledge the funding support from the EPSRC/MRC-funded Neuromod+ and Closed-loop Neural Interface Technologies (closeNIT) Network+. AG acknowledges support from the Royal Commission for the Exhibition of 1851 and the L'Oréal-UNESCO UK and Ireland For Women in Science Rising Talent Programme. TRM was supported by the Wellcome Trust (215575/Z/19/Z) and the Medical Research Council (MC_UU_00030/10). TC is funded by the NIHR Newcastle Biomedical Research Centre awarded to the Newcastle upon Tyne Hospitals NHS Foundation Trust, Newcastle University and Cumbria, Northumberland, Tyne and Wear Foundation Trust. The authors would like to extend our sincere gratitude to Dr Karen Rommelfanger (Institute of Neuroethics), and Dr Antonio Valentin (King's College London), Neil Roberts (Science and Engineering Health Technologies Alliance, SEHTA), Rebecca Woodcock (NIHR MindTech MedTech Co-operative), Vanessa Pinfold (The McPin Foundation), Annabel Walsh (The McPin Foundation), Prof Sian Robinson (University of Newcastle) and Chris Gibbs (Cumbria, Northumberland, Tyne and Wear NHS Foundation Trust) for their valuable feedback on earlier versions of this manuscript. Their insights and suggestions were instrumental in refining the content and enhancing the overall quality of our work. For the purpose of open access, the author has applied a Creative Commons Attribution (CC BY) licence to any Author Accepted Manuscript version arising from this submission.

**DATA AVAILABILITY STATEMENT**
No data has been generated in this study.

**REFERENCES**
1. Federal Register: Methods and Leading Practices for Advancing Public Participation and Community Engagement With the Federal Government. https://www.federalregister.gov/documents/2024/03/20/2024-05882/methods-and-leading-practices-for-advancing-public-participation-and-community-engagement-with-the.
2. Smith, H. *et al.* Co-production practice and future research priorities in United Kingdom-funded applied health research: a scoping review. *Heal. Res. Policy Syst.* **20**, 1–43 (2022).
3. Munce, S. E. *et al.* Development of the Preferred Components for Co-Design in Research Guideline and Checklist: Protocol for a Scoping Review and a Modified Delphi Process. *JMIR*

*Res. Protoc.* **12**, e50463 (2023).
4. Starke, G. *et al.* Qualitative studies involving users of clinical neurotechnology: a scoping review. *BMC Med. Ethics* **25**, 1–14 (2024).
5. Price, A. *et al.* Frequency of reporting on patient and public involvement (PPI) in research studies published in a general medical journal: a descriptive study. *BMJ Open* **8**, e020452 (2018).
6. Holmes, L. *et al.* Innovating public engagement and patient involvement through strategic collaboration and practice. *Res. Involv. Engagem.* **5**, 1–12 (2019).
7. Abelson, J. *et al.* PUBLIC AND PATIENT INVOLVEMENT IN HEALTH TECHNOLOGY ASSESSMENT: A FRAMEWORK FOR ACTION. *Int. J. Technol. Assess. Health Care* **32**, 256–264 (2016).
8. Stilgoe, J., Lock, S. J. & Wilsdon, J. Why should we promote public engagement with science? *https://doi.org/10.1177/0963662513518154* **23**, 4–15 (2014).
9. Phillips, O. R. *et al.* What are the strengths and limitations to utilising creative methods in public and patient involvement in health and social care research? A qualitative systematic review. *Res. Involv. Engagem.* **10**, 1–13 (2024).
10. Goering, S. *et al.* Recommendations for Responsible Development and Application of Neurotechnologies. *Neuroethics* **14**, 365–386 (2021).
11. Robinson, J. T. *et al.* Building a culture of responsible neurotech: Neuroethics as socio-technical challenges. *Neuron* **110**, 2057–2062 (2022).
12. The Royal Society. *IHuman Blurring Lines between Mind and Machine*. https://royalsociety.org/-/media/policy/projects/ihuman/report-neural-interfaces.pdf (2019).
13. Ramos, K. M. & Koroshetz, W. J. Integrating ethics into neurotechnology research and development: The US National Institutes of Health BRAIN Initiative®. *Neuroethics Anticip. Futur.* 144–156 (2017) doi:10.1093/OSO/9780198786832.003.0008.
14. Wood, C. R., Xi, Y., Yang, W. J. & Wang, H. Insight into Neuroethical Considerations of the Newly Emerging Technologies and Techniques of the Global Brain Initiatives. *Neurosci. Bull.* **39**, 685–689 (2023).
15. Jwa, A. S. & Poldrack, R. A. Addressing privacy risk in neuroscience data: from data protection to harm prevention. *J. Law Biosci.* **9**, 1–25 (2022).
16. UNESCO. *Unveiling the Neurotechnology Landscape. Scientific Advancements Innovations and Major Trends*. *Unveiling the neurotechnology landscape. Scientific advancements innovations and major trends* (2023) doi:10.54678/OCBM4164.
17. van der Scheer, L., Garcia, E., van der Laan, A. L., van der Burg, S. & Boenink, M. The Benefits of Patient Involvement for Translational Research. *Heal. Care Anal.* **25**, 225–241 (2017).
18. Sacristán, J. A. *et al.* Patient involvement in clinical research: why, when, and how. *Patient Prefer. Adherence* **10**, 631–640 (2016).
19. Patient and public involvement and engagement in research - a 'how to' guide for researchers. Doi:10.5281/zenodo.3515811.
20. UK Standards for Public Involvement. https://sites.google.com/nihr.ac.uk/pi-standards/home.
21. Public engagement – UKRI. https://www.ukri.org/what-we-do/public-engagement/.
22. Embedding lived experience in mental health research | Funding Guidance | Wellcome. https://wellcome.org/grant-funding/guidance/embedding-lived-experience-expertise-mental-health-research#what-if-i'm-working-on-a-basic-science-project?-63b9.
23. European Commission. Horizon Europe - European Commission. https://research-and-innovation.ec.europa.eu/funding/funding-opportunities/funding-programmes-and-open-calls/horizon-europe_en.
24. RRI Tools. Public Engagement - RRI Tools. https://rri-tools.eu/public-engagement.
25. Canadian Institutes of Health Research. CIHR's Framework for Citizen Engagement - CIHR. https://cihr-irsc.gc.ca/e/41270.html.
26. National Health and Medical Research Council. Consumer and community engagement | NHMRC. https://www.nhmrc.gov.au/about-us/consumer-and-community-

involvement/consumer-and-community-engagement.
27. National Institutes of Health. Public Involvement with NIH | National Institutes of Health (NIH). https://www.nih.gov/about-nih/what-we-do/get-involved-nih/public-involvement-nih.
28. Cluley, V. *et al.* Mapping the role of patient and public involvement during the different stages of healthcare innovation: A scoping review. *Heal. Expect.* **25**, 840–855 (2022).
29. World Economic Forum. *Empowering 8 Billion Minds Enabling Better Mental Health for All via the Ethical Adoption of Technologies*. www.weforum.org (2019).
30. PPIE Planner. https://plan4ppie.com/.
31. Science, Society and Engagement – An e-Anthology is now published | Engage2020. https://engage2020.eu/news/science-society-and-engagement-an-e-anthology-is-now-published/.
32. ActionCatalogue - methods. http://actioncatalogue.eu/search.
33. Public Engagement Roadmap. https://pe-roadmap.azureedge.net/.
34. Definition and role of the designated PPI (Patient and Public Involvement) lead in a research team | NIHR. https://www.nihr.ac.uk/documents/definition-and-role-of-the-designated-ppi-patient-and-public-involvement-lead-in-a-research-team/23441.
35. Consumer and Community Involvement - AHRA. https://ahra.org.au/our-work/consumer-and-community-involvement/.
36. Eugene Washington PCORI Engagement Award Program | PCORI. https://www.pcori.org/engagement-research/eugene-washington-pcori-engagement-award-program.
37. Schedule of Activities 2022-2027 - Cambridge Biomedical Research Centre. https://cambridgebrc.nihr.ac.uk/public/patient-and-public-involvement-engagement-and-participation-ppie-strategy-2022-2027/ppie-strategy-2022-2027/.
38. Funding Opportunities | NCCPE. https://www.publicengagement.ac.uk/funding-opportunities.
39. Overview | Shared decision making | Guidance | NICE. https://www.nice.org.uk/guidance/ng197.
40. Involvement Cost Calculator | INVOLVE. https://www.invo.org.uk/resource-centre/payment-and-recognition-for-public-involvement/involvement-cost-calculator/.
41. Payment for public involvement in health and care research: a guide for organisations on employment status and tax - Health Research Authority. https://www.hra.nhs.uk/planning-and-improving-research/best-practice/public-involvement/resources/payment-public-involvement-health-and-care-research-guide-organisations-employment-status-and-tax/.
42. Considerations when paying patient partners in research - CIHR. https://cihr-irsc.gc.ca/e/51466.html.
43. Tools and resources | George & Fay Yee Centre for Healthcare Innovation | University of Manitoba. https://umanitoba.ca/centre-for-healthcare-innovation/tools-and-resources.
44. Health Equity Assessment Tool (HEAT): executive summary - GOV.UK. https://www.gov.uk/government/publications/health-equity-assessment-tool-heat/health-equity-assessment-tool-heat-executive-summary.
45. Member directory | Association of Medical Research Charities. https://www.amrc.org.uk/Pages/Category/member-directory?Take=20.
46. Medical research organizations | Cause IQ. https://www.causeiq.com/directory/medical-research-organizations-list/.
47. Jinks, C. *et al.* Patient and public involvement in primary care research-an example of ensuring its sustainability. (2016) doi:10.1186/s40900-016-0015-1.
48. Witham, M. D. *et al. Developing a Roadmap to Improve Trial Delivery for Under-Served Groups: Results from a UK Multi-Stakeholder Process*. *Trials* vol. 21 https://www.nihr.ac.uk/documents/improving-inclusion-of-under-served-groups-in-clinical-research-guidance-from-include-project/25435 (2020).

49. Equality Impact Assessment (EqIA) Toolkit | arc-em.nihr.ac.uk. https://arc-em.nihr.ac.uk/clahrcs-store/equality-impact-assessment-eqia-toolkit.
50. Health Determinants Research Collaborations | NIHR. https://www.nihr.ac.uk/explore-nihr/support/health-determinants-research-collaborations.htm.
51. Education That Empowers - EUPATI. https://eupati.eu/.
52. SPOR Evidence Alliance. https://sporevidencealliance.ca/.
53. Consumer and community representatives in peer review for Targeted Calls for Research | NHMRC. https://www.nhmrc.gov.au/about-us/consumer-and-community-involvement/consumer-and-community-representative-involvement-peer-review-process-targeted-calls-research.
54. Health Nexus Santé. https://healthnexus.ca/.
55. Consumers Health Forum of Australia | Consumers Shaping Health. https://chf.org.au/.
56. Swasti | Home. https://swasti.org/.
57. thebigword. https://en-gb.thebigword.com/.
58. Patient and public involvement impact log – NIHR Health Protection Research Unit in Behavioural Science and Evaluation at University of Bristol. https://hprubse.nihr.ac.uk/public-involvement/patient-and-public-involvement-impact-log/.
59. Brett, J. *et al.* Reaching consensus on reporting patient and public involvement (PPI) in research: methods and lessons learned from the development of reporting guidelines. *BMJ Open* **7**, e016948 (2017).
60. Staniszewska, S. *et al.* GRIPP2 reporting checklists: tools to improve reporting of patient and public involvement in research. *BMJ* **358**, 3453 (2017).
61. Cook, N., Siddiqi, N., Twiddy, M. & Kenyon, R. Patient and public involvement in health research in low and middle-income countries: a systematic review. *BMJ Open* **9**, e026514 (2019).
62. Shields, G. E., Brown, L., Wells, A., Capobianco, L. & Vass, C. Utilising Patient and Public Involvement in Stated Preference Research in Health: Learning from the Existing Literature and a Case Study. *Patient* **14**, 399–412 (2021).
63. Bequette, B. W. Challenges and recent progress in the development of a closed-loop artificial pancreas. *Annu. Rev. Control* **36**, 255–266 (2012).